\documentstyle[nato,epsf]{crckapb} 

\begin{opening}
\title{NUCLEATION THEORY OF MAGNETIZATION SWITCHING \protect\\
       IN NANOSCALE FERROMAGNETS}

\author{Per Arne Rikvold$^{1,2,3}$}
\author{M.A.\ Novotny$^2$}
\author{M.~Kolesik$^2$}
\institute{$^1$Center for Materials Research and Technology, 
           $^2$Supercomputer Computations Research Institute, 
           and $^3$Department of Physics\\
           Florida State University\\
           Tallahassee, FL 32306-3016, USA}
\author{Howard L.\ Richards}
\institute{Max-Planck-Institut f{\"u}r Polymerforschung\\
D-55128 Mainz, Germany}

\end{opening}

\runningtitle{NUCLEATION THEORY OF SWITCHING}

\begin{document}


\begin{abstract}
A nucleation picture of 
magnetization switching in single-domain ferromagnetic nanoparticles 
with high local anisotropy is discussed. 
Relevant aspects of nucleation theory are presented, stressing 
the effects of the particle size on the switching dynamics. 
The theory is illustrated by Monte Carlo simulations and compared 
with experiments on single particles. 
\end{abstract}

\section{Introduction}
\label{sec-intro}

The dynamics of magnetization switching in nanometer-sized particles of 
highly anisotropic ferromagnets is interesting, from both the 
scientific and technological points of view. 
The basic scientist sees in such particles a laboratory to study the 
decay of a metastable phase towards equilibrium, 
while the technologist sees a promising material for 
ultrahigh-density magnetic recording media. 
Although ferromagnetic nanoparticles have been studied experimentally 
for a long time \cite{KNEL63}, until recently this was only possible 
with powders. However, with modern techniques of 
nanofabrication \cite{KENT93} 
and ultrahigh-resolution methods to detect the magnetization, 
such as Magnetic Force Microscopy (MFM) \cite{CHAN93} 
Lorentz microscopy \cite{SALL91} 
and micro-SQUID devices 
\cite{WERN}, 
one can now synthesize and study such particles individually. 

The most common description of magnetization switching is a 
mean-field approach, originally due to N{\'e}el \cite{NEEL49} 
and Brown \cite{BROWN}. To avoid an energy barrier due to 
exchange interactions of strength $J$, 
uniform rotation of all the atomic moments in the particle is assumed. The 
remaining energy barrier, $\Delta$, is caused by magnetic anisotropy, which  
is a combination of crystal-field and magnetostatic effects. 
The equilibrium thickness of a wall between 
oppositely magnetized domains is $\xi \propto \sqrt{J/\Delta}$. For particles 
smaller than $\xi$, the uniform-rotation picture is reasonable.  
If the anisotropy is largely magnetostatic, 
the resulting demagnetizing field causes particles larger than $\xi$ 
to form oppositely magnetized domains, and 
switching is achieved through the field-driven 
motion of preexisting domain walls. 
However, if the anisotropy is largely due to the 
local crystalline environment, there exists a window of particle sizes 
that are larger than $\xi$ but 
smaller than the size at which the particle becomes multidomain. 
[This is for instance often the case in ultrathin films.] 
Such particles can be modeled as Ising systems with local 
spin variables, $s_i = \pm 1$. Depending on the degree of anisotropy, 
these spins can either represent the $z$ component of 
individual atomic moments, or 
one can coarse grain the system by rescaling all lengths in terms of 
$\xi$, so that the $s_i$ represent block spins. 
The Ising Hamiltonian is 
\begin{equation}
\label{eq:ham}
{\cal H}_0 = - J \sum_{\langle i,j \rangle} s_i s_j - H \sum_i s_i \;.
\end{equation}
Here $J > 0$ is the ferromagnetic exchange interaction, $H$ is the applied 
magnetic field times the local magnetic moment, and the sums 
$\sum_{\langle i,j \rangle}$ and $\sum_i$ run over all nearest-neighbor 
pairs and all sites on a suitable lattice, respectively. 
Here we only report numerical results for two-dimensional square lattices, 
but our theoretical arguments are valid for general spatial dimension. 
The order parameter is the dimensionless magnetization, 
\begin{equation}
\label{eq:mag}
m = N^{-1} \sum_i s_i \;,
\end{equation}
where $N$ is the total number of Ising spins in the particle. 

In the highly anisotropic nanoscale 
ferromagnets described by Eq.~(\ref{eq:ham}) [and modifications 
discussed below], the state of uniform magnetization opposite to 
the applied field is properly viewed as a {\em metastable phase\/}. 
This nonequilibrium phase decays by 
neither uniform rotation nor by the motion of preexisting 
domain walls, but rather by the thermal 
nucleation and subsequent growth of {\em localized droplets}, 
inside which the magnetization is parallel with the field \cite{RICH94}. 
This decay mechanism yields results very similar to 
effects observed in recent experiments on well-characterized 
single-domain 
ferromagnets in the nanometer range. In this paper we concentrate on 
a maximum in the switching field (or coercivity) vs.\ particle 
size \cite{CHAN93}. 
Another quantity which is often measured in experiments,
is the probability that the particle has {\em not\/} switched within a 
specified waiting time \cite{SALL91,WERN}. Results 
concerning this quantity 
can be found in Refs.~\cite{RICH94} and~\cite{NOVO97}. 

The Ising model does not have an intrinsic dynamic. To simulate the effects of 
thermal fluctuations we therefore use a local stochastic dynamic 
which does not conserve the order parameter, 
such as the ones proposed by Metropolis {\em et al.\/} \cite{METR53} 
or Glauber \cite{GLAU63}. In order to perform simulations on the 
very long timescales necessary to observe metastable decay, we use a 
so-called ``rejection-free" Monte Carlo (MC) algorithm \cite{NOVOTNY}. 
The basic time scale of the MC simulation [MC Steps per Spin (MCSS)] 
is not known from first principles and must be fitted to experiments. 
It is expected to be on the order of a typical inverse 
phonon frequency, 10$^{-9}$--10$^{-13}$~s.

\section{Nucleation and Growth}
\label{sec-NG}

This section is a brief primer on the theory of nucleation and growth 
as it applies to systems in the dynamic universality class of 
kinetic Ising models with nonconserved order parameter. 
{}For further details, see
Refs.~\cite{RICH94,RICH95C,RICH96,TOMITA,RIKV94}. 

\subsection{Background} 
\label{sec-NG1}

The central problems in nucleation theory are to identify the 
fluctuations that lead to the decay of the metastable phase and to 
obtain their free-energy cost, relative to the 
metastable phase. For Ising-like systems with short-range interactions, 
these fluctuations are compact droplets of radius $R$. The magnetization 
inside the droplet is parallel with the applied field and has a 
magnitude near 
the temperature dependent 
zero-field magnetization, 
$m_{\rm sp}(T)$, which is nonzero below the critical temperature, $T_c$. 
The free energy of the droplet has two competing 
terms: a positive surface term $\propto R^{d-1}$, and a 
negative bulk term $\propto |H| R^d$, where $d$ is the spatial dimension.
The competition between these terms yields a critical droplet radius, 
\begin{equation}
  \label{eq:Rc}
        R_c(H,T) = \frac{(d-1)\sigma(T)}
                       {2 |H| m_{\rm sp}(T)} \; ,
\end{equation}
where $\sigma(T)$ is the surface tension. Droplets with $R < R_c$ 
most likely decay, whereas droplets with $R > R_c$ most likely grow further 
to complete the switching process. The free-energy cost of the 
critical droplet ($R = R_c$) is 
\begin{equation}
   \label{eq:DFd-pbc}
        \Delta F_{\rm SD}(H,T) = \Omega_d \sigma(T)^d
        \left( \frac{d \! - \! 1}{2 |H| m_{\rm sp}(T)} \right)^{d-1} ,
\end{equation}
where $\Omega_d$ is a weakly $T$ dependent 
shape factor such that the volume of a 
droplet of radius $R$ equals $\Omega_d R^d$. 
The subscript SD stands for Single Droplet, as explained below. 
Nucleation is a stochastic process, and the nucleation rate per unit volume 
is given by a Van't Hoff-Arrhenius relation: 
\begin{equation} \label{eq:NucRate}
        I(H,T) \propto 
|H|^{K} \exp \left[ - \frac{\Delta F_{\rm SD}(H,T) }{k_{\rm B}T} \right] 
\equiv 
|H|^{K} \exp \left[ - \frac{\Xi(T)}{k_{\rm B}T |H|^{d-1}} \right] 
\; ,
\end{equation}
where $k_{\rm B}$ is Boltzmann's constant, and $\Xi(T)$ is the $H$-independent 
part of $\Delta F_{\rm SD}$. 
The prefactor exponent $K$ 
equals 3 for the two-dimensional Ising model
and $-1/3$ for the three-dimensional Ising model 
\cite{RIKV94,LANGER,GNW80}.

\subsection{Effects of Finite Particle Size} 
\label{sec-NG2}

{}For particles of finite linear size, $L$, an important crossover occurs 
for combinations of $H$, $T$, and $L$, such that $R_c \approx L$. 
This yields a $T$ and $L$ dependent crossover field 
called the Thermodynamic Spinodal (ThSp) \cite{TOMITA,RIKV94}: 
\begin{equation}
\label{eq:HThSp}
H_{\rm ThSp}(T,L) \approx \frac{(d-1)\sigma(T)}
                       {2 m_{\rm sp}(T) L} \; .
\end{equation}

{}For $|H| < H_{\rm ThSp}$, $R_c$ would exceed $L$. 
This is called the Coexistence (CE) regime 
because the critical fluctuation in such weak fields 
resembles two coexisting slabs of opposite magnetization 
\cite{TOMITA,RIKV94}. 
The average metastable lifetime in the CE regime is approximately 
\begin{equation}
\label{eq:tauCE}
\tau_{\rm CE}(H,T,L) \sim 
\exp \left[ \frac{2 \sigma (T) L^{d-1} - 2 A m_{\rm sp}(T) |H| L^d}
                 {k_{\rm B} T} \right] \; ,
\end{equation}
where $A$ is a nonuniversal constant.
Since $|H| \le H_{\rm ThSp} \sim L^{-1}$, the dominant size dependence is 
an exponential increase with $L^{d-1}$. 
This behavior also holds for more general 
boundary conditions than the periodic boundary conditions used to 
obtain Eq.~(\ref{eq:tauCE}) \cite{RICH96}. 

{}For $|H| > H_{\rm ThSp}$ (but not too large, as we shall see below), 
the lifetime is dominated by the inverse of the total nucleation rate, 
\begin{equation}
\label{eq:tauSD}
\tau_{\rm SD}(H,T,L) \approx \left( L^d I(H,T) \right)^{-1} 
\propto 
L^{-d} |H|^{K} \exp \left[ \frac{\Xi(T)}{k_{\rm B}T |H|^{d-1}} \right] 
\; .
\end{equation}
It is inversely proportional to the particle volume, $L^d$. 
The subscript SD stands for Single Droplet  
and indicates that in this regime the switching is completed by the first 
droplet whose radius exceeds $R_c$. 

A second crossover, called the Dynamic Spinodal (DSp) 
\cite{TOMITA,RIKV94}, 
is predicted when one observes that a supercritical 
droplet grows at a finite velocity, which for large 
droplets is proportional to the field: $v \approx \nu |H|$. 
A reasonable criterion to locate the DSp is that the average time between 
nucleation events, $\tau_{\rm SD}$, should equal the time 
it takes a droplet to grow to a size comparable to $L$. This leads to the 
asymptotic relation 
\begin{equation}
\label{eq:HDSP}
        H_{\rm DSp}(T,L)  \sim
        \frac{(d-1)}{2 m_{\rm sp}(T)}
        \left[ \frac{\Omega_d \sigma(T)^d}
                {(d+1) k_{\rm B} T \ln L } \right]^{\frac{1}{d-1}} .
\end{equation}
{}For $|H| > H_{\rm DSp}$, the metastable phase decays through many droplets 
which nucleate and grow independently in different parts of the system. 
This is called the Multidroplet (MD) regime 
\cite{TOMITA,RIKV94}. A classical theory of metastable decay in 
large systems \cite{KOLM37,JOHN39,AVRAMI} 
gives the lifetime in this regime,
\begin{equation}
\label{eq:tauMD}
\tau_{\rm MD}(H,T) 
\approx \left[ \frac{I(H,T) \Omega_d (\nu |H|)^d}{(d+1) \ln 2}
            \right]^{- \frac{1}{d+1}}
\;,
\end{equation}
which is {\em independent\/} of $L$. 

The switching field, $H_{\rm sw}(t_{\rm w},T,L)$, is the field 
required to observe a specified average lifetime, $t_{\rm w}$. 
It is found by solving 
Eqs.~(\ref{eq:tauCE}), (\ref{eq:tauSD}), and~(\ref{eq:tauMD}) for $H$ 
with $t_{\rm w}$ for the respective average lifetimes, $\tau_{\rm CE}$, etc.
The resulting $L$ dependence of $H_{\rm sw}$ 
is illustrated by the MC data shown in 
Fig.~\ref{fig:JMMM}(a). It consists of a steep increase with $L$ 
in the CE regime, 
peaking at the ThSp, followed by a decrease in the SD regime towards a 
plateau in the MD regime.

\section{Numerical Results}
\label{sec-NR}

In this section we present some representative 
results of simulations of two-dimensional Ising ferromagnets, 
which we compare with the theoretical predictions of the previous section 
and with experiments.
 
\subsection{Pure System with Periodic Boundary Conditions}
\label{sec-2DI}

\begin{figure}
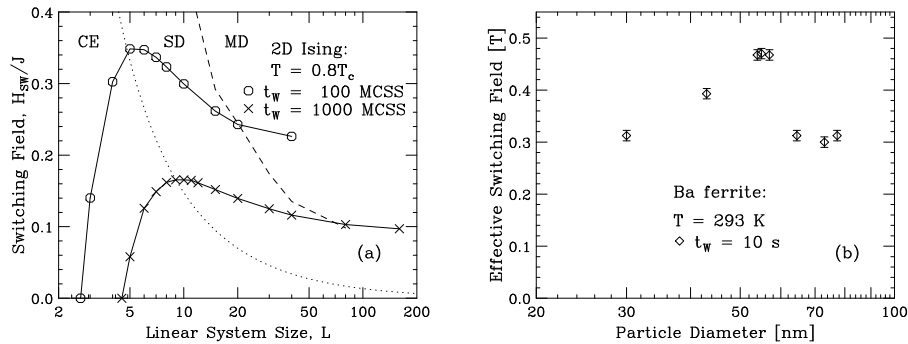

\vspace{4.5truecm}

 \includegraphics{JMMM2c.psc}

 \includegraphics{judy5.psc}

\caption[]{
Switching fields vs.\ particle size. 
(a):
MC simulations for a two-dimensional Ising ferromagnet with periodic 
boundary conditions. The dotted line is the ThSp, 
and the dashed line is the DSp. 
After Ref.~\protect\cite{RICH94}. 
(b):
Effective switching fields for nanoscale Ba-ferrite particles. 
Data digitized from Fig.~5 of Ref.~\protect\cite{CHAN93}. 
}
\label{fig:JMMM}
\end{figure}

The simplest model considered is a two-dimensional 
square-lattice Ising system with periodic boundary conditions. The 
switching fields for this model at $T = 0.8 \, T_c$ 
are shown in Fig.~\ref{fig:JMMM}(a) for $t_{\rm w} = 100$ and 1000~MCSS. 
The $L$ and $t_{\rm w}$ dependencies 
expected from the results of Sec.~\ref{sec-NG2} are clearly seen. 
We emphasize that the decrease in the SD region is {\it not\/} due to 
an equilibrium domain structure.
It is an entropy effect of purely dynamical origin, 
arising from the volume factor in Eq.~(\ref{eq:tauSD}) \cite{RICH96}. 
Analogous corrections to nucleation rates in fluids 
were proposed by Lothe and Pound \cite{LOTH62}. 

For qualitative comparison we show in Fig.~\ref{fig:JMMM}(b) 
effective switching fields for nanoscale 
Ba-ferrite particles, obtained by MFM experiments \cite{CHAN93}. 
We propose that the peak observed in the switching field may be of the 
same purely dynamical origin as in kinetic Ising models. 

\subsection{Effects of a Demagnetizing Field}
\label{sec-Demag}

\begin{figure}
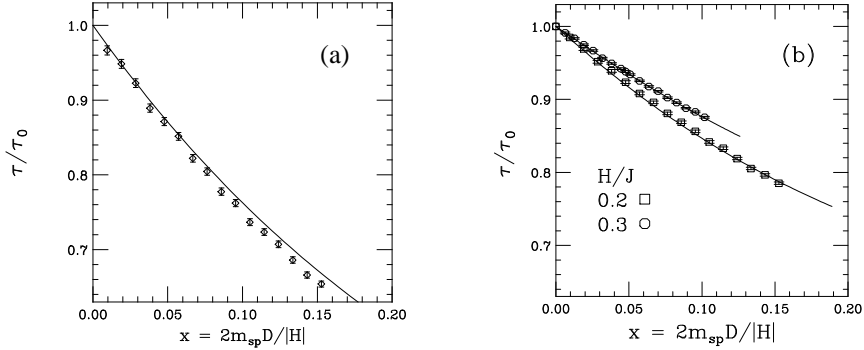

\vspace{4.5truecm}

 \includegraphics{sd_tau_dx.psc}

 \includegraphics{md_tau_dx.psc}

\caption[]{
Relative changes in 
the average metastable lifetime versus the reduced demagnetizing 
factor, $x = 2 D m_{\rm sp}(T) / |H|$, at $T = 0.8T_c$. 
The solid curves are analytical 
results that only require parameters determined for $D = 0$. 
After Ref.~\protect\cite{RICH95C}. 
(a):
SD regime, $|H| = 0.2J$, $L = 10$. 
(b):
MD regime, $L = 100$. 
}
\label{fig:demag}
\end{figure}

A reasonable objection to the model defined by 
Eq.~(\ref{eq:ham}) is the absence of dipolar interactions, 
which causes it to be single domain for all $L$. 
To address this shortcoming without the large 
computational expense of recalculating dipole sums at every step 
in the dynamical simulation, a model was introduced in which 
the demagnetizing field was approximated by adding 
a weak long-range antiferromagnetic term: 
${\cal H}_D = {\cal H}_0 + D L^d m^2$ \cite{RICH95C}. 
Particles smaller than $L_D \approx 2 \sigma(T) / D $ remain single 
domain \cite{KITT46}, 
but the demagnetizing factor $D$ decreases the free-energy barrier 
towards nucleation of the equilibrium phase. 
Addition of the demagnetizing factor 
was found to reduce the average lifetime by an analytically 
predictable amount, as shown in Fig.~\ref{fig:demag}. 
However, no qualitative 
differences from the behavior described above were observed. 

\subsection{Heterogeneous Nucleation}
\label{sec-hetero}

Next we discuss ways in which the homogeneous nucleation 
observed in pure systems with periodic boundary conditions is modified by 
heterogeneous nucleation at the particle surface or at quenched 
inhomogeneities. 

\subsubsection{Modified Boundary Conditions}
\label{sec-FBC}

The use of periodic boundary conditions allows one to study 
bulk nucleation without complications due to the particle surface. 
Since the surface can be modified in various ways by 
reconstruction, adsorption, oxidation, etc., one cannot in general predict 
whether it will enhance or inhibit nucleation. However, even 
addition to the Ising model of a surface field or modified 
exchange interactions at the surface 
produces complicated crossovers between surface and bulk nucleation  
\cite{RICH96}. In general, the changes   
reduce the height of the peak in $H_{\rm sw}$ vs.\ $L$, but 
for a wide range of modifications it remains clearly discernible. 
Examples are shown in Fig.~\ref{fig:hetero}(a). 

\subsubsection{Quenched Randomness}
\label{sec-QR}

\begin{figure}
\vspace{4.5truecm}

 \includegraphics{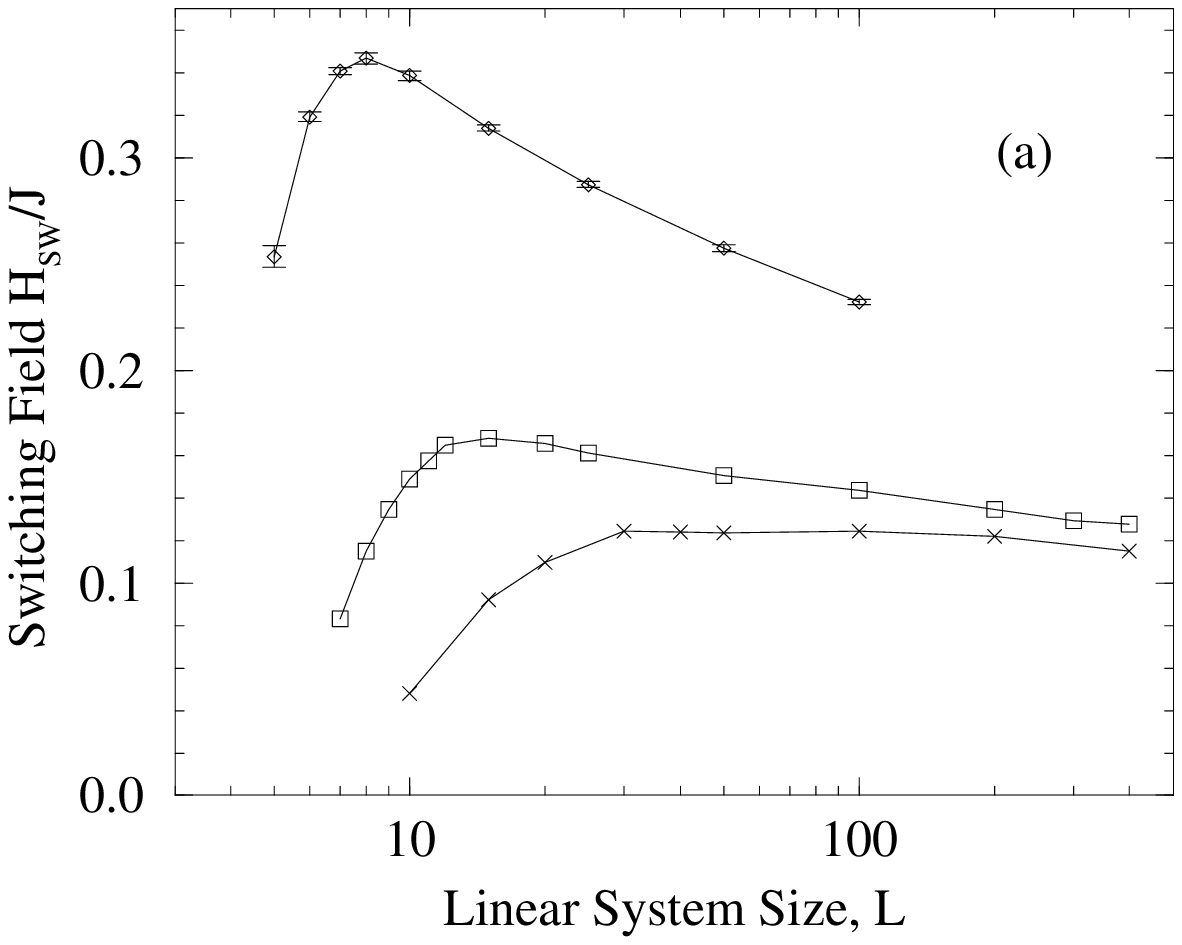}

 \includegraphics{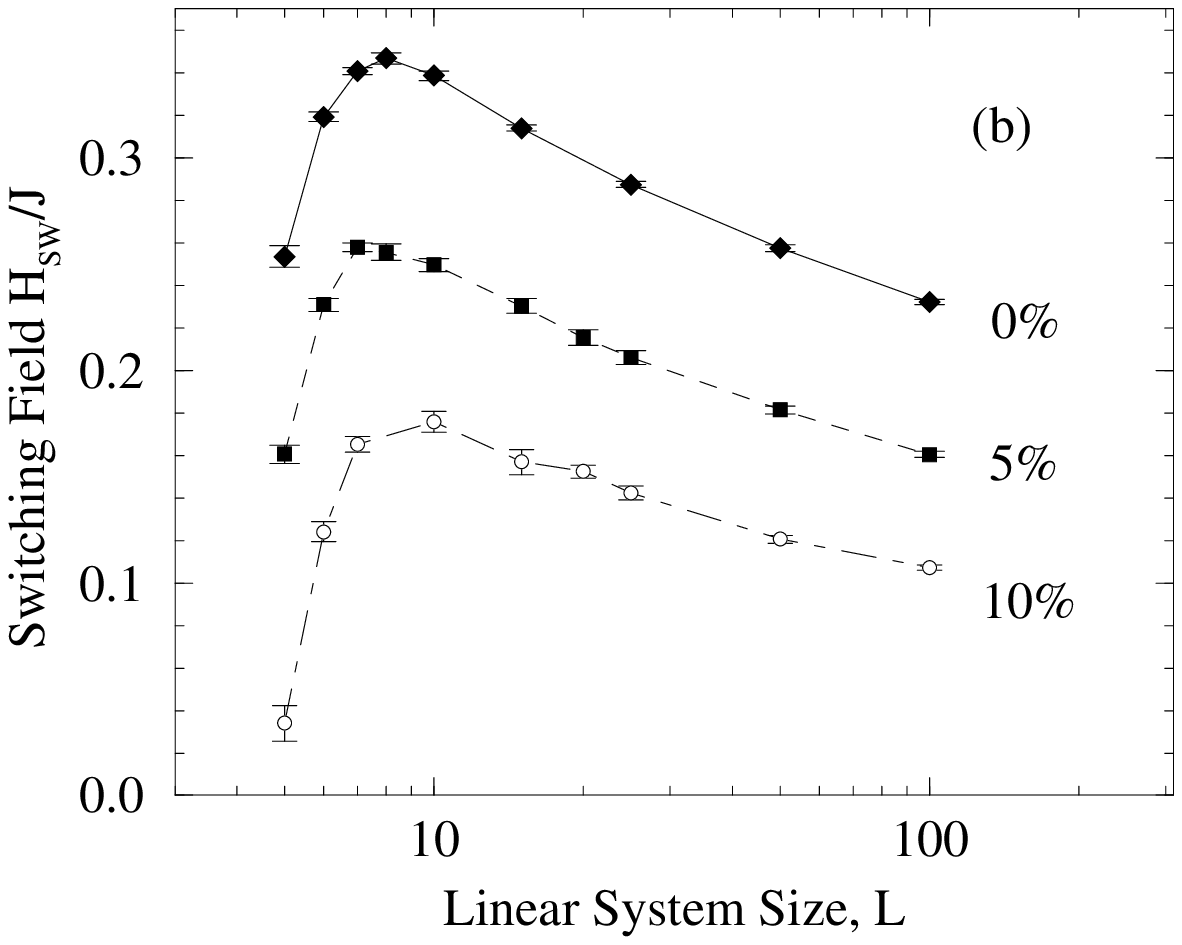}

\caption[]{
Effects of heterogeneous nucleation on $H_{\rm sw}$ at $T \approx 0.57\,T_c$ 
with $t_{\rm w} = 30000$~MCSS. For comparison, 
the top curve in both panels corresponds to homogeneous nucleation in 
a pure system with periodic boundary conditions. 
(a):
Effects of boundary conditions in a pure system. 
Middle curve: square system with periodic boundary conditions 
in one direction and free boundary conditions in the other. 
Bottom curve: circular system with free boundary conditions. 
Data from Ref.~\protect\cite{RICH96}. 
(b):
Effects of random bond dilution in a system with 
periodic boundary conditions. 
After Ref.~\cite{KOLE96}. 
}
\label{fig:hetero}
\end{figure}

Another way in which heterogeneous nucleation may dominate,
is through quenched impurities. An exploratory study 
was presented in Ref.~\cite{KOLE96}. 
Bond dilution was observed to reduce $H_{\rm sw}$ by a factor approximately 
independent of $L$, as shown in Fig.~\ref{fig:hetero}(b), while random 
spin magnitudes led to non-self-averaging behavior and a wide distribution of 
lifetimes. 

\subsubsection{Coercivity of Fe Sesquilayers on W(110)}
\label{sec-FeW}

Much interest has recently been devoted to ultrathin iron films 
on W(110) substrates \cite{ELME90,ELME94,BETH95,SAND96,SKOM96}. 
The so-called sesquilayer systems, 
which consist of islands of a second monolayer of Fe on top of an almost 
perfect first monolayer \cite{BETH95}, have particularly interesting 
magnetic properties \cite{SAND96,SKOM96,SUEN97}. 
Around an Fe coverage of approximately 1.5~monolayers (ML), the coercivity 
exceeds that of a monolayer or a bilayer by more than an order of 
magnitude \cite{SKOM96}. 

Magnetization switching in this system is expected to occur through the 
field-driven motion of preexisting domain walls, which are pinned at the 
second-layer islands. Based on this picture, the coercivity has been 
calculated by micromagnetic methods \cite{SAND96,SKOM96}. However, those 
calculations did not consider thermal effects and 
were also essentially static. 

To account for thermal depinning and the dependence of the 
coercivity on the frequency of the applied field, a two-layer Ising model 
has been developed for this system \cite{KOLE97A}. 
This is a reasonable approximation 
since the crystal-field anisotropy for Fe monolayers on W(110) 
is almost two orders 
of magnitude larger than for bulk Fe \cite{ELME90}.
Simulations were performed on a computational lattice in which the 
second-layer island morphology was reproduced using STM images of real 
systems \cite{BETH95}, and the exchange 
interactions were chosen to reproduce the experimentally observed 
critical temperature of an Fe monolayer, $T_c = 230$~K \cite{ELME94}. 
\begin{figure}
\vspace{8.0truecm}

 \includegraphics{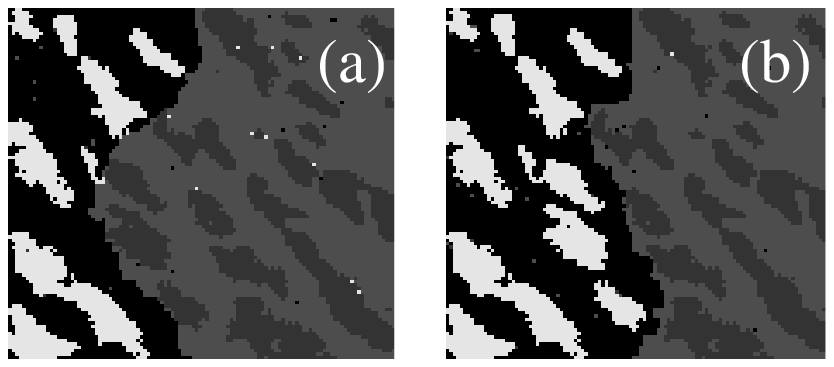}

 \includegraphics{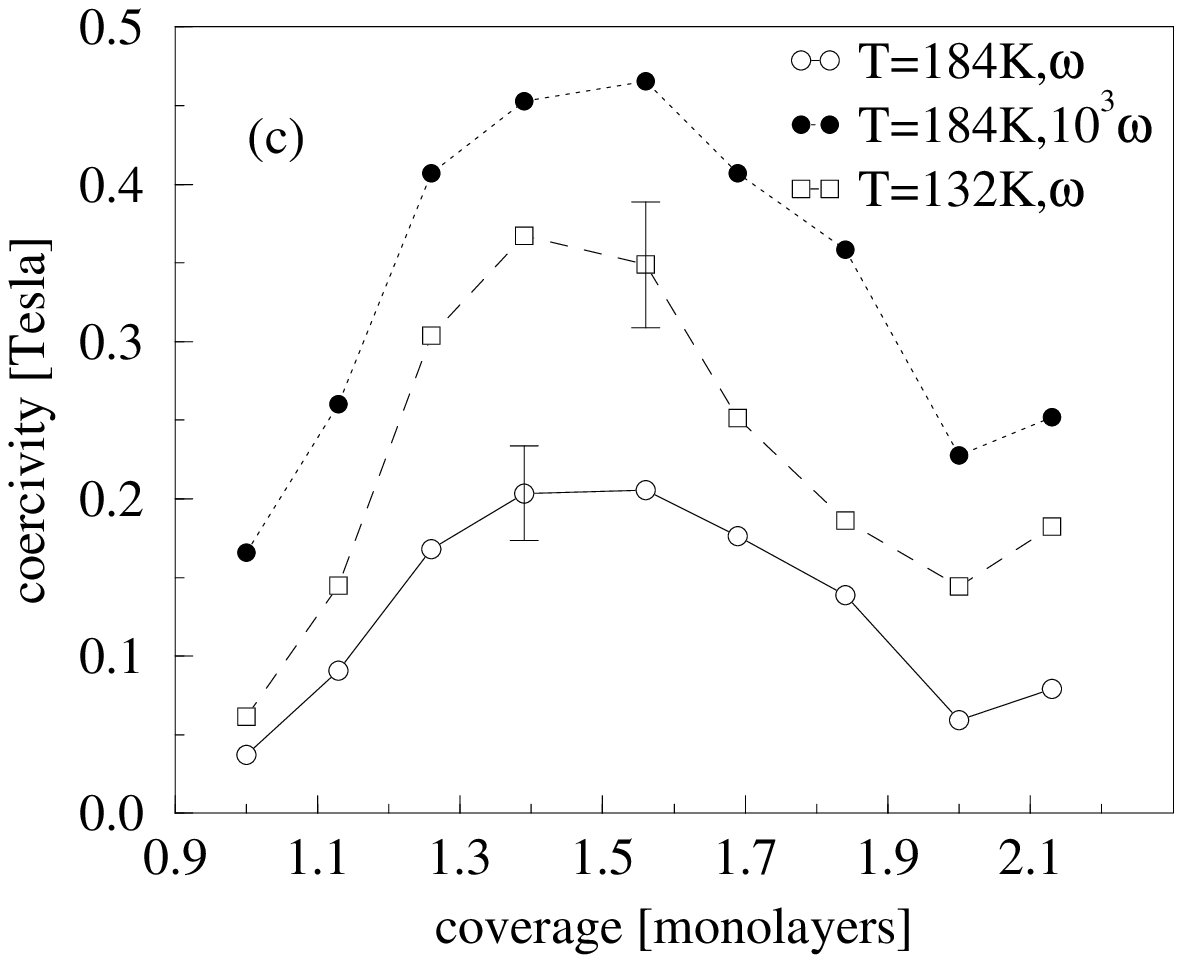}

 \includegraphics{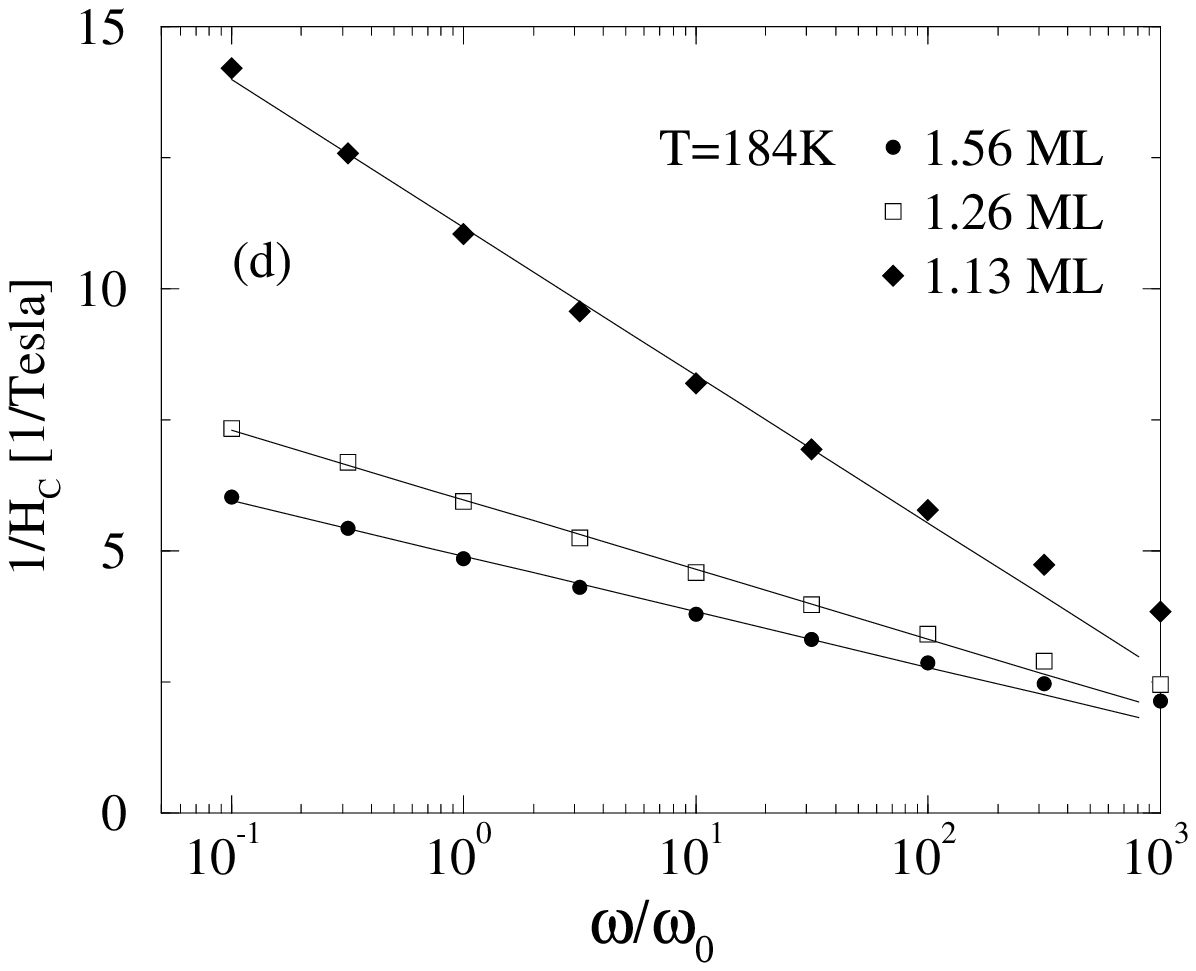}

\caption[]{
Simulations of magnetization switching in Fe sesquilayers on W(110). 
After Ref.~\cite{KOLE97A}. 
(a) and (b): 
Snapshots of a domain wall propagating across a sesquilayer 
with coverage 1.26~ML.
The high-contrast region represents the growing equilibrium phase. 
The area shown is 654~{\AA} $\times$ 610~{\AA} 
[109 $\times$ 102 computational cells], and the island configuration was 
digitized from Fig.~1~j) of Ref.~\protect\cite{BETH95}. 
The simulated temperature and field correspond to 132~K and 0.26~T, 
respectively. The time elapsed between the two snapshots is 
approximately 1.5$\times 10^{6}$~MCSS, 
corresponding to 1.5$\times 10^{-6}$~s. 
A movie of this simulation is found at 
{\tt http://www.scri.fsu.edu/$\sim$rikvold}.
(c):
Sesquilayer coercivity vs.\ Fe coverage, estimated by extrapolation 
to weak fields.
The lower curve should be 
compared with Fig.~3 of Ref.~\protect\cite{SKOM96}. 
(d):
The frequency dependence of the estimated coercivity. 
}
\label{fig:FeW}
\end{figure}

Two simulation snapshots of the magnetic domain wall moving across a 
sesquilayer are shown in Figs.~\ref{fig:FeW}(a) and~(b). The wall moves 
intermittently, spending most of its time pinned 
against the ``windward" side of 
the islands. Thermally nucleated depinning events are followed by rapid 
advances to the next metastable position. The coercivity is estimated 
from the average domain-wall velocity.
Estimated coercivities for two different temperatures and 
driving frequencies are shown vs.\ the Fe coverage in Fig.~\ref{fig:FeW}(c). 
The experimentally observed nonmonotonic coverage dependence \cite{SKOM96} 
is reproduced, as well as the temperature \cite{SAND96} 
and frequency \cite{SUEN97} dependencies. 
The model yields an approximately linear 
dependence of the inverse coercivity on the logarithm of the 
frequency, shown in Fig.~\ref{fig:FeW}(d). 
Over a few decades of frequency, this is hard to distinguish numerically 
from a power law \cite{SUEN97}.

\section{Conclusions} 
\label{sec-Con}

We have presented a brief overview of a nucleation theory of magnetization 
switching in single-domain ferromagnets in the nanometer range. We emphasized 
the dependence of the switching field or coercivity on the particle size 
and demonstrated that the model is capable of reproducing the 
experimentally observed maximum in the switching field vs.\ particle size.

Our discussion places the switching dynamics of nanoscale ferromagnets 
in the context of metastable decay in finite systems. 
This interdisciplinary field is experiencing a renaissance due to new
methods of nanofabrication and observation of individual
systems. In addition to magnets, 
results have recently been published for systems as
different as liquid mixtures \cite{LALA97}
and semiconductor nanocrystals \cite{CHEN97}.

\section*{Acknowledgements}

Supported in part by Florida State University through the 
Center for Materials Research and Technology and the  
Supercomputer Computations Research Institute 
(Department of Energy Contract DE-FC05-85ER25000), 
and by National Science Foundation Grants DMR-9520325 and DMR-9315969. 
Computing resources at the
National Energy Research Supercomputer Center were provided by
the Department of Energy.


\end{document}